\documentclass[prl,aps,epsf,twocolumn,showpacs,10pt,superscriptaddress ]{revtex4-1}

\usepackage{graphicx,amsfonts,times,bm,amsmath,verbatim,color,array}
\newcommand{\<}{\langle}
\newcommand{\e}{\varepsilon}
\renewcommand{\>}{\rangle}

\usepackage{natbib}
\usepackage{graphicx,amssymb,amsmath,ifthen,braket,xcolor}
\usepackage[linktocpage=true,colorlinks=true,urlcolor=blue,linkcolor=red,citecolor=blue]{hyperref}
\usepackage{bm}

\usepackage{xr}
\makeatletter
\newcommand*{\addFileDependency}[1]{
  \typeout{(#1)}
  \@addtofilelist{#1}
  \IfFileExists{#1}{}{\typeout{No file #1.}}
}
\makeatother

\newcommand*{\myexternaldocument}[1]{
    \externaldocument{#1}
    \addFileDependency{#1.tex}
    \addFileDependency{#1.aux}
}
\myexternaldocument{SM_2}
\begin{document}

% \title{Nonlinear Planar Hall effect: tilt induced spin-to-charge conversion }
\title{Quantum oscillations as a robust fingerprint of chiral anomaly in nonlinear response in Weyl semimetals}
%\title{Quantum oscillations of the nonlinear planar effects signifying chiral anomaly in Weyl Semimetals}
\date{\today}

\author{Chuanchang Zeng}
\affiliation{Centre for Quantum Physics, Key Laboratory of Advanced Optoelectronic Quantum Architecture and Measurement(MOE),
School of Physics, Beijing Institute of Technology, Beijing, 100081, China}
\affiliation{Beijing Key Lab of Nanophotonics $\&$ Ultrafine Optoelectronic Systems,
School of Physics, Beijing Institute of Technology, Beijing, 100081, China}

\author{Snehasish Nandy}
\affiliation{Department of Physics, University of Virginia, Charlottesville, VA 22904, USA}

\author{Pu Liu}
\affiliation{Centre for Quantum Physics, Key Laboratory of Advanced Optoelectronic Quantum Architecture and Measurement(MOE),
School of Physics, Beijing Institute of Technology, Beijing, 100081, China}
\affiliation{Beijing Key Lab of Nanophotonics $\&$ Ultrafine Optoelectronic Systems,
School of Physics, Beijing Institute of Technology, Beijing, 100081, China}

\author{Sumanta Tewari}
\affiliation{Department of Physics and Astronomy, Clemson University, Clemson, SC 29634, USA
}

\author{Yugui Yao}
\thanks{ygyao@bit.edu.cn}
\affiliation{Centre for Quantum Physics, Key Laboratory of Advanced Optoelectronic Quantum Architecture and Measurement(MOE),
School of Physics, Beijing Institute of Technology, Beijing, 100081, China}
\affiliation{Beijing Key Lab of Nanophotonics $\&$ Ultrafine Optoelectronic Systems,
School of Physics, Beijing Institute of Technology, Beijing, 100081, China}

\begin{abstract}
%In view of searching for the signature of the celebrated chiral anomaly (CA) in Weyl semimetals (WSMs) in ongoing experiments, quantum oscillation in the linear response regime has been considered an important fingerprint in the magneto transports in WSMs.
We investigate the nonlinear planar effects (NPEs) in Weyl semimetals (WSMs) starting from the semiclassical regime to the ultra-quantum limit within the framework of Boltzmann transport theory incorporating Landau quantization. Based on our results, we propose quantum oscillations in the NPEs as a robust signature of the celebrated chiral anomaly (CA) in WSMs. By obtaining analytical expressions, we show that the quantum oscillations in the non-linear regime exhibit two different periods in $B^{-1}$ ($B$ is the magnetic field), compared to the linear response regime with only one period in the inverse magnetic field. 
%We find that the quantum oscillations in NPEs are attributed to the deviation in the chiral chemical potential, and therefore directly linked to CA. 
In addition, we obtain characteristic angular dependence of the CA-induced NPEs. We conclude that in light of the inconclusive sign of the CA-driven longitudinal magneto-conductance in WSMs as has been illustrated in recent theoretical work, the proposed behaviors of quantum oscillations in the non-linear planar effects uniquely identify the existence of chiral anomaly in WSMs.
\end{abstract} 
%\pacs{}

\maketitle %\section{Introduction}\paragraph{\textcolor{blue}{Introduction.\textemdash{}}}
\textcolor{blue}{\it{Introduction:}} Weyl semimetals (WSMs) have attracted great interest theoretically and experimentally in recent years for offering a plethora of intriguing physical phenomena. Besides the spectroscopic observations of the characteristic Weyl nodes and Fermi arc states, detecting various quantum effects intrinsic to Weyl fermions has also been vastly employed in the identification of the WSMs. Chiral anomaly (CA) or Adler–Bell–Jackiw anomaly is one such intrinsic effect unique to WSMs~\cite{Ninomiya_ca_1983plb}. In the presence of a non-orthogonal electric and magnetic field i.e. $\bm{E\cdot B}\neq 0$, CA pumps charge between Weyl nodes of opposite chirality. This inter-node flow of charge, balanced with inter-node scattering with characteristic time $\tau_v$, creates a finite steady-state chiral chemical potential (CCP) $\mu^{C}$ proportional to $\tau_v\bm{E\cdot B}$, leading to fascinating transport effects. Typical examples are the positive longitudinal magneto-conductivity (LMC)~\cite{spivak_LMC_2013prb,burkov_magnetotransport_2014prb} along with the planar Hall effect (PHE)~\cite{Burkov_phe_2017prb,Nandy_phe_2017prl}.
%, which is the transverse version of LMC. 
Either in the semiclassical regime for small magnetic fields or in the quantum limit with large magnetic fields involving Landau levels, it has been found that the CA-induced longitudinal  magneto-conductivity in WSMs is positive and increases with magnetic field $\bm{B}$ quadratically and linearly, respectively. This renders the measurements of longitudinal magneto-conductivity (i.e., $\bm{B}||\bm{J}$, with $\bm{J}$ the current density),  typically small in regular metals because of the absence of Lorentz forces, an important signature of chiral anomaly in WSMs~\cite{He_expLMC_2014prl_DSM,liang_expLMC_2015nm_DSM,li_expLMC_2016np_DSM,zhang_expLMC_2016nc,hirschberger_expLMC_2016nm_SM,li_expLMC_2017fop,zhang_expLMC_2017nanoLett, Kumar_phe_2018prb, Chen_phe2018prb, Liang_currentjet_2018prx,jianhuiZhou_2018PRB,songboZhang_2020PRB, Tanay_2022PRB, Zhang_2016NJP, Johannes_2019PRB, Johannes_2016PRB}.

Interestingly, it has recently been reported that not only the LMC and PHE in WSMs can enhance through various extrinsic mechanisms~\cite{Arnold_currentjet_2016Nc,Dos_currentjet_2016NJPh,Liang_currentjet_2018prx, Yang-currentjet_2019prb,li_expLMC_2017fop,yamada_negativemass_2021} other than CA, but also, they can acquire a negative sign as well, either in the semiclassical ~\cite{Knoll_negativeLMC_2020prb_weakB,Xiao_negativeLMC_2020prb_weakB,Girish_negative_2021prb_weakB1,Girish_negative_2021prb_weakB2} or quantum regime~\cite{Lu_negativeLMC_2015prb_strongB, Xie_negativeLMC_2016prb_strongB,Sankar_negativeLMC_2016prb_strongB}. Refs.~\cite{Wiedmann_LMC_2015prb_TIs,Dai_LMC_2017prl_TIs,Assaf_LMC_2017prl_TIs,Spivak_LMC_2018prl_convSM,Chang_PHE_FM_2019PRB,Yin_PHE_AFM_2019prl,Groen_PHE_FM_2021pra,Li_phe_crys_2010jpcm,Seemann_phe_crys_2011prl,Taskin_phe_2017nc_TIs,Nandy_phe_2018sr_TIs,Zheng_phe_2020prb_TIs,Archit_phe_2021Apl_TIs} show that the positive LMC and PHE can even appear in materials without any Weyl nodes. As a result, the positive LMC and PHE can no longer be considered sufficient transport signatures for CA, and alternative approaches to identifying the existence of CA in WSMs are highly required~\cite{Liang_reviewpaper_2021}.

\begin{figure}[tp!]
	\begin{center}
		\includegraphics[width=0.465\textwidth]{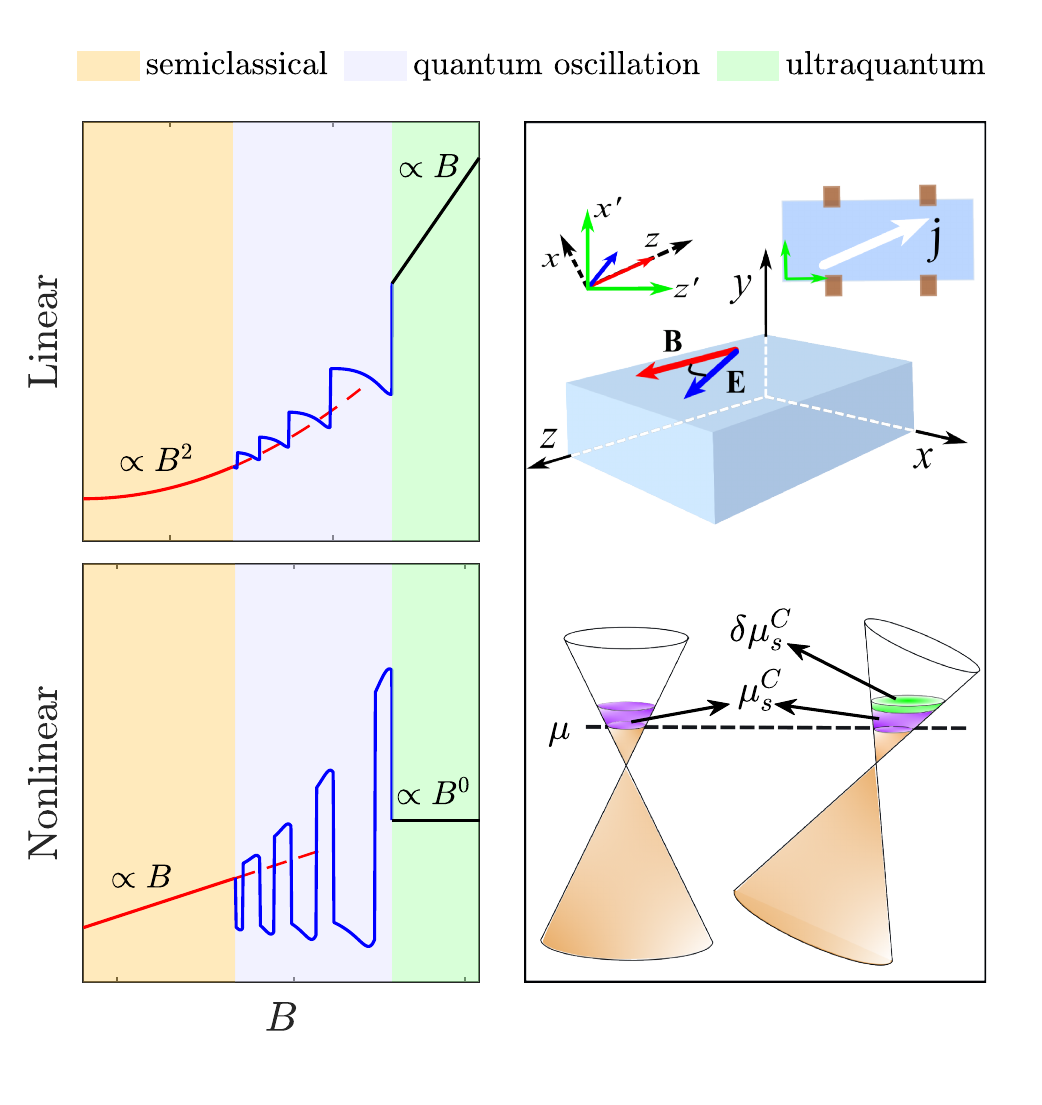}
		\llap{\parbox[b]{163mm}{\large\textbf{(a)}\\\rule{0ex}{75mm}}}
		\llap{\parbox[b]{164mm}{\large\textbf{(b)}\\\rule{0ex}{40mm}}}
		\llap{\parbox[b]{78mm}{\large\textbf{(c1)}\\\rule{0ex}{74mm}}}
		\llap{\parbox[b]{78mm}{\large\textbf{(c2)}\\\rule{0ex}{40mm}}}
	\end{center}
	\vspace{-11mm}
	\caption{(Color online) The phenomenological magnetic field dependence of the linear and nonlinear magneto planar conductivity induced by chiral anomaly (CA) in WSMs are shown in (a) and (b) respectively, for three transport regimes. Panel (c1) presents the geometry of a crossed fields lying in the $x-z$-plane, and the schematic setup for experimental probing (\textit{top right}, inset) along with the possible coordinate systems (\textit{top left}, inset). (d) shows the CA-induced chiral chemical potential (purple, $\mu^C_s$) and its deviation (green, $\delta \mu^C_s$) for a single Weyl cone. } \label{fig:fig1}
	\vspace{-2mm}
\end{figure}

Recent work proposed that a periodic-in-$B^{-1}$ quantum oscillation in the transport coefficients, crossing the semiclassical and ultra-quantum limits  in the linear response regime, is a unique fingerprint of CA~\cite{Deng_qc_2019prl,osci_exp_zhang_2016_prb,osci_exp_du_2016_scp}. 
%Interestingly, such quantum oscillations, which inherently arise from the CA-induced chiral chemical potential and are not plagued by other extrinsic effects, have been observed in recent experiments~\cite{osci_exp_zhang_2016_prb,osci_exp_du_2016_scp}. 
By contrast, in the nonlinear response regime, it has been shown that the transport signatures can be understood as a combined effect of CA and Berry curvature-induced anomalous velocity~\cite{Li_caNLH_2021prb,Nandy_caNLH_2021prb,Zeng_caNTH_2021} or even purely as a Berry curvature effect~\cite{Zeng_caNTH_2021} within the semiclassical framework. 
%Although these nonlinear responses have been proposed to be resulting from CA-induced transport in WSMs, it has been found that similar behavior can also originate merely from the Berry curvature effect in the semiclassical limit~\cite{Zeng_caNTH_2021}. 
Therefore, a convincing and unambiguous connection between magneto-transport and CA in the nonlinear regime for WSMs is still lacking. 
In this Letter, we propose a new way to detect CA in WSMs under the action of a moderate magnetic field, via the quantum oscillation in the nonlinear response of LMC and PHE, namely, the \textit{nonlinear} planar effects (NPEs). Using  Boltzmann transport theory that incorporates Landau quantization, we find that a unique feature of the quantum oscillations of the NPEs is the existence of two different oscillation periods in $B^{-1}$, one of which remains constant while the other decreases with increasing $B^{-1}$. We show that these NPE responses rely on the deviations of chiral chemical potential i.e., $\delta \mu^C$, and are directly linked to CA. The quantum oscillation behavior of the NPEs can be contrasted with the CA-driven linear planar effects, 
%which are ascribed to the CCP $\mu^C$ %in the presence of $\bm{E\cdot B \neq 0}$, 
which oscillate with only one period in $B^{-1}$. We also obtain the $B$-dependent NPE responses in the weak-$B$ limit, consistent with previous semiclassical results~\cite{Li_caNLH_2021prb}. Interestingly, going beyond the semiclassical regime, we find that the NPEs at the ultra-quantum limit are magnetic field-independent. The comparison of the magnetic field dependencies of CA-induced magneto-transport between the linear ~\cite{spivak_LMC_2013prb,burkov_magnetotransport_2014prb, Ninomiya_ca_1983plb,spivak_LMC_2013prb,ultraq_burkov_2012prb,osci_Shovkov_2014prb,Deng_qc_2019prl,kamal_cas_2020prr} and the nonlinear regime is illustrated in Fig.~\ref{fig:fig1}.
Based on these calculations, we conclude that, in light of  the inconclusive sign of the CA-driven longitudinal magneto-conductance in WSMs \cite{Knoll_negativeLMC_2020prb_weakB,Xiao_negativeLMC_2020prb_weakB,Girish_negative_2021prb_weakB1,Girish_negative_2021prb_weakB2,Lu_negativeLMC_2015prb_strongB, Xie_negativeLMC_2016prb_strongB,Sankar_negativeLMC_2016prb_strongB}, and the existence of such behavior even in the absence of chiral anomaly \cite{Wiedmann_LMC_2015prb_TIs,Dai_LMC_2017prl_TIs,Assaf_LMC_2017prl_TIs,Spivak_LMC_2018prl_convSM,Chang_PHE_FM_2019PRB,Yin_PHE_AFM_2019prl,Groen_PHE_FM_2021pra,Li_phe_crys_2010jpcm,Seemann_phe_crys_2011prl,Taskin_phe_2017nc_TIs,Nandy_phe_2018sr_TIs,Zheng_phe_2020prb_TIs,Archit_phe_2021Apl_TIs},  the proposed behaviors of quantum oscillations in the non-linear planar effects coupled with similar oscillations in the linear response regime, uniquely identify the existence of chiral anomaly in WSMs.

%%%%%%%%%%%%%%%%%%%%%%%%%%%%%%%%%%%%%%%%%%%%%%%%%%%%%%%%%
%%%%%%%%%%%%%%%%%%%%%%%%%%%%%%%%%%%%%%%%%%%%%%%%%%% FIG 1
% here we do not discuss the DOS, instead we leave it in SM and mention it later
\textcolor{blue}{\it Landau levels  of tilted WSMs:} The low-energy effective Hamiltonian describing a pair of tilted Weyl nodes can generally be written as,
\begin{equation}
    H(\bm{k})= \sum_s s v_F \hbar \bm{k}\cdot \bm{\sigma} +s( \hbar \bm{w}\cdot \bm{k} + Q_0)\sigma_0 \label{eq: hamiltonian_0}
\end{equation}
where $\sigma_{0}$ and $\bm{\sigma}=(\sigma_{x},\sigma_{y},\sigma_{z})$ represent Pauli matrices, $s=\pm 1$ is the chirality for each node, $v_F$ denotes the isotropic Fermi velocity in the absence of tilt, and vector $\bm{w}$ tilts the Weyl node along axis-$i$ with strength $w_{i}~(i=x, y, z)$. Without loss of generality, a band tilt lying within the $yz$-plane i.e., $\bm{w}=(0, w_y, w_z)$ is considered for our discussion hereinafter~\cite{Tchoumakov_lorentzbst_2016prl}. A finite energy $Q_0$, which has been shown earlier~\cite{Li_caNLH_2021prb,Nandy_caNLH_2021prb} to be important in generating the nonlinear Hall response induced by CA is also included here. 

\begin{figure}[tp!]
	\begin{center}
		\includegraphics[width=0.465\textwidth]{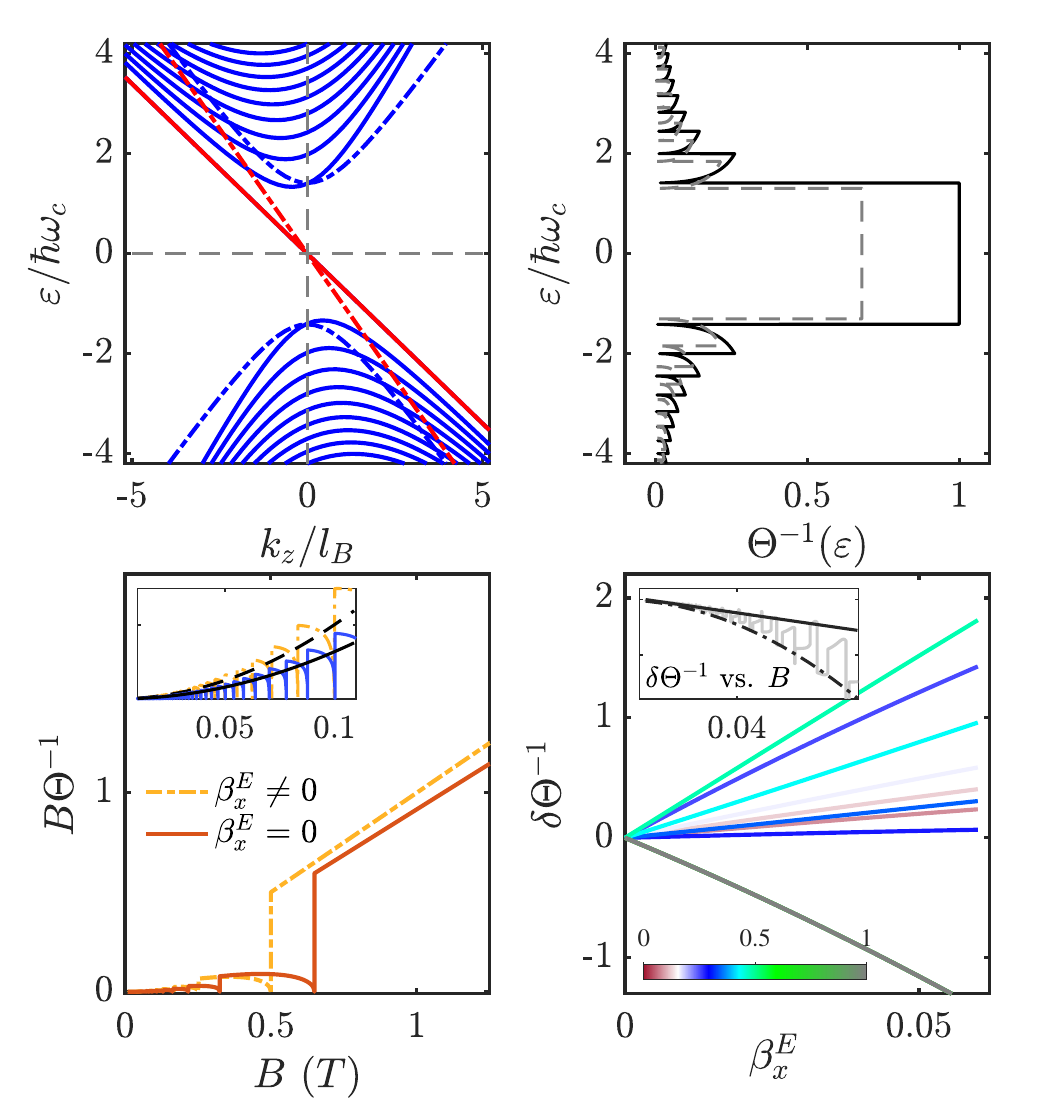}
		\llap{\parbox[b]{162mm}{\large\textbf{(a)}\\\rule{0ex}{81mm}}}
		\llap{\parbox[b]{84mm}{\large\textbf{(b)}\\\rule{0ex}{80mm}}}
		\llap{\parbox[b]{162mm}{\large\textbf{(c)}\\\rule{0ex}{42mm}}}
		\llap{\parbox[b]{84mm}{\large\textbf{(d)}\\\rule{0ex}{42mm}}}
	\end{center}
	\vspace{-7mm}
	\caption{(Color online) (a) Landau levels (LLs) projected in the $k_y=0$ plane for Weyl node $s=+1$, where the solid (dashed) lines represent the LLs of a tilted (non-tilted) Weyl node. Panel (b) presents the oscillating DOSs $\Theta^{-1}$ as function of Fermi energy. The dashed (solid) curves imply for the case with ${\beta s \neq 0}$ i.e. ${\beta^E_x=0.1, \beta^w_y=0.1}$ (${\beta s=0}$, i.e. ${\beta^E_x=\beta^w_y=0}$). To better reveal the modification, here a finite $w_z =0.3$ is taken for the case with $\beta_s \neq 0$. Panel (c) plots $B\Theta^{-1}$ as function of $B$, where the inset shows the quadratic-in-B dependence in the weak-B limit. (d) depicts the linear dependency of $\delta \Theta^{-1}$ on $\beta^E_x$ at various magnetic field strengths (indicated by the color scale). The inset shows $\delta \Theta^{-1} \propto B$ in the weak-B limit.
	} \label{fig:fig2}
	\vspace{-2mm}
\end{figure}

We proceed by introducing the electromagnetic fields represented by $\mathcal{A}=(\phi, \bm{A})$ with the vector potential in Landau gauge $\bm{A}=(0, xB, -E_{z} t)$ and the scalar potential $\phi = -xE_{x}$, which yield $\bm{B}=\bm{\nabla\times A} = B\hat{\bm{z}}$ and $\bm{E}=-\nabla \phi -\partial_t \bm{A} =(E_{x}, 0, E_{z})$~\cite{Deng_qc_2019prb}, respectively. Such a configuration is shown in Fig.~\ref{fig:fig1}(c1). %To preserve the translational invariance along $z$-direction (or keep the momentum $k_z$ be a good quantum number), a time-dependent vector potential $A_z = E_{z} t$ is applied here to introduce the parallel component of the electric field~\cite{Deng_qc_2019prb}. 
Following the standard Peierls substitution $\bm{k}\rightarrow \bm{q} =\bm{k}+e\bm{A}/\hbar$, Eq.~(\ref{eq: hamiltonian_0}) can be rewritten as,
\begin{equation}
    H(\bm{q})= \sum_s s v_F \hbar \bm{q}\cdot \bm{\sigma} +s(\hbar \bm{w}\cdot\bm{q} + Q_0)\sigma_0 + e x E_{x}. \label{eq:hamiltonian_EB}
\end{equation}
It is clear from the Eq.~(\ref{eq:hamiltonian_EB}) that the band tilt generates an effective electric field $E_{eff} \hat{\bm{x}} = s w_y B\hat{\bm{x}}$ in addition to the external field $E_x\hat{\bm{x}}$. Now to obtain the Landau levels (LLs), we consider a Lorentz boost along $y$-axis (perpendicular to $\bm{E}$ and $\bm{B}$) in terms of the relativistic parameter $\beta$ with 
\begin{equation}
    %\tanh{{\vartheta}} =
    \beta=\beta^{w}_{y} +\beta^{E}_{x}  \label{eq:beta}
\end{equation} 
where $\beta^{w}_{y}=s w_{y} /v_F$ and $\beta^{E}_{x} =E_{x}/v_FB$.  
%provided by band tilt and the external electric field, respectively.
Applying this boost transformation and its inverse operation successively, the LLs for Eq.~(\ref{eq:hamiltonian_EB}) can be obtained as $\e^{s}_{n}(k_y, k_z) =  s (Q^{\prime}_0+ w_{z}\hbar k_z) + s_0\alpha \sqrt{(v_F\hbar k_z)^2 + 2\alpha |n| (\hbar \omega_c)^2}$,  
\begin{comment}
\begin{equation}\begin{split}
     \e^{s}_{n}(k_y, k_z) = & s (Q^{\prime}_0+ w_{z}\hbar k_z) + \\
     & s_0\alpha \sqrt{(v_F\hbar k_z)^2 + 2\alpha |n| (\hbar \omega_c)^2}  \label{eq:LLs_n}
 \end{split}
\end{equation}\end{comment}
with $s_0=\text{sign}(n)=\pm 1$ indicating the $n^{th}$ positive or negative LLs for $|n|\geq 1$. Here $\omega_c =v_F/l_B$ with $l_{B} = \sqrt{\hbar/eB}$ is defined as the cyclotron frequency, characterizing the LL spacing. Factor $\alpha=\sqrt{1-\beta^2}$ indicates the squeezing effect on the LLs as well as the cyclotron frequency. The finite energy $Q_0$ is now modulated as $Q^{\prime}_0 = Q_0 -s\beta^{E}_{x} \hbar k_y$, which incorporates a $k_{y}$-dependent band shift brought about by the Lorentz transformation. 
%also arises due to the Lorentz transformation, which provides a transverse drift velocity for each LL~\cite{Deng_qc_2019prb} and is now incorporated into the modulated $Q^{\prime}_0 = Q_0 -s\beta^{E}_{x} \hbar k_y$. 
The zeroth-LL ($n=0$), similarly, is found as $\e^{s}_{0}(k_y, k_z) = s (Q^{\prime}_0+ w_{z}k_z -\alpha v_F\hbar k_z) \label{eq:LLs_0}$. With the band dispersion in hand, the longitudinal group velocity $v^s_{z,n}$ for the LLs can be obtained straightforwardly~\cite{SM}. The LL dispersion for both the tilted and non-tilted Weyl node with $s=+1$ is depicted in Fig.~\ref{fig:fig2}(a). Obviously, the zeroth-LL dispersing to the right is chiral (in red) while the higher ones are achiral (in blue). 
%%%%%%%%%%%%%%%%%%%%%%%%%%%%%%%%%%%%%%%%%%%%%%%%%%%%%%%%%
%%%%%%%%%%%%%%%%%%%%%%%%%%%%%%%%%%%%%%%%%%%%%%%%%%% FIG 2

It is important to note that, the motion of the electrons governed by Eq.~(\ref{eq:hamiltonian_EB}) can be arranged into quantized LLs only within the so-called magnetic regime~\cite{classicalEM,classicalFields}, as revealed in previous works either merely when $\beta^{E}_{x}<1$~\cite{Lukose_lorentz_graphene_2007prl,Peres_Lorentz_graphene_2007iop,Arjona_lorentz_wsm_2017prb,Deng_qc_2019prl,Deng_qc_2019prr} or $\beta^{\omega}_y <1$~\cite{Yu_lorentz_mageto_2016prl,Tchoumakov_lorentzbst_2016prl,Shao_lorentz_wsm_2021iop}. In contrast, here $|\beta|=|\beta^{w}_{y}+\beta^{E}_{x}| <1 $ is expected to support the quantization. This also implies a more general collapse/breakdown of LLs, which, in our case, is no longer restricted to type-II WSMs~\cite{Yu_lorentz_mageto_2016prl,Tchoumakov_lorentzbst_2016prl}. As it now depends on the joint effects of the external electric field and band tilt, and hence the LL-collapse can appear even for type-I WSMs.  
%competition/cooperation between the electric field and band tilt. 

\textcolor{blue}{\it{Boltzmann formalism decorated with Landau quantization:}} For a single Weyl node with chirality $s$, the phenomenological Boltzmann transport equation is given by $\big[ \frac{\partial}{\partial t} + \bm{\dot{r}}^{s}\cdot \bm{\nabla}_{\bm{r}} +\bm{\dot{k}}^{s}\cdot \bm{\nabla}_{\bm{k}} \big] f^{s}_{\bm{k},n} = \mathcal{I}_{coll}(f^{s}_{\bm{k},n})$
\begin{comment}
\begin{equation}
    \bigg[ \frac{\partial}{\partial t} + \bm{\dot{r}}^{s}\cdot \bm{\nabla}_{\bm{r}} +\bm{\dot{k}}^{s}\cdot \bm{\nabla}_{\bm{k}} \bigg] f^{s}_{\bm{k},n} = \mathcal{I}_{coll}(f^{s}_{\bm{k},n}) \label{eq:bte}
\end{equation}\end{comment}
where $f^{s}_{\bm{k},n}$ represents the non-equilibrium Fermi-Dirac distribution function for electrons with energy $\e^{s}_{\bm{k},n}$. We consider the relaxation time approximation for the collision integral $\mathcal{I}_{coll}(f^{s}_{\bm{k},n})$ that involves both the intranode ($\tau_a$) and internode ($\tau_v$) scattering processes in WSMs. Specifically, the scattering time $\tau_a$ relaxes the electron from $f^{s}_{\bm{k},n}$ to a local equilibrium state $f^{s}_{eq}$, while $\tau_v$ indicates the relaxation to a global equilibrium state $f^{0}_{eq}$. It is worthy to note that we have $\langle f^{s}_{\bm{k},n}\rangle_{s} = f^{s}_{eq}$, with $\langle {\mathcal{ }}\rangle_{s}$ averaging over all the possible electron states~\cite{Deng_qc_2019prl,Zeng_caNTH_2021}.
 
Solving the Boltzmann equation in the current setup~\cite{SM}, the energy deviation $\delta \e^{s}_{\bm{k},n}$ perturbed by fields can be obtained as $\delta \e^{s}_{\bm{k},n}= -\tau_a e\bm{E}\cdot \bm{v}^{s}_{\bm{k},n}
       + (1 -\tau_{a}/\tau_{v} )\mu^{C}_s$, with $\mu^{C}_s$ given as
\begin{equation}
    \mu^{C}_s =\mu_{s}-\mu = -\tau_v e \langle\bm{E\cdot v^{s}_{\bm{k},n}}\rangle_{s}, \label{eq:deltamu} % \langle \delta \e^{s}_{\bm{k},n}\rangle_{s}=
\end{equation}
exactly the CA-induced CCP for the Weyl node with chirality $s$. %Immediately, the perturbed distribution function is found as $f^{s}_{\bm{k},n} = f^0_{eq} +\big[-\partial_{\e_{\bm{k}}} f^0_{eq}(\bm{k})\big] \delta \e^{s}_{\bm{k},n} \label{eq:FD_expansion}$, and thus the charge current is $\bm{j}= \sum_{s, n,k_y} \int \frac{dk_z}{2\pi} (-e)\bm{v}^{s}_{\bm{k},n}f^{s}_{\bm{k},n}$. 
In the geometry of $\bm{E\cdot B}\neq0 $ lying in the $x-z$-plane, the longitudinal ($\parallel$ with $\bm{B}$) charge current is found to be $\bm{j}_{z} =\bm{j}^{C}_{z}+\bm{j}^{N}_{z}$, i.e. composed of the CA-induced response $\bm{j}^{C}_{z}$ and the normal drift response $\bm{j}^{N}_{z}$, respectively written as
\begin{equation}\begin{split}
     \bm{j}^{C}_{z} = c_0 v_F(\tau_{v}-\tau_{a})\frac{E_{z}B}{\mathcal{F}_{\Theta}}, 
     \quad \bm{j}^{N}_{z} = c_0 v_F\tau_a \mathcal{F}_{\Lambda} E_{z}B \label{eq:currents}
\end{split}\end{equation}
\begin{comment}\begin{equation}\begin{split}
     \bm{j}^{C}_{z} &= c_0 v_F(\tau_{v}-\tau_{a})  E_{z}B/\mathcal{F}_{\Theta} , \\
     \bm{j}^{N}_{z} &= c_0 v_F\tau_a \mathcal{F}_{\Lambda} E_{z}B \label{eq:currents}
\end{split}\end{equation}\end{comment}
with $c_0 = 2  e^3/h^2$. The auxiliary function $\mathcal{F}_{x}$ here is defined as $\mathcal{F}_{x} =\int d\e [-\partial_{\e} f_{0}(\e)] x (\e) \label{eq:Fs}$, with $x=\Theta(\e)$, $\Lambda(\e)$ (see the details in SM~\cite{SM}). In the zero-temperature limit where the quantized LLs are not smeared from temperature broadening effect~\cite{SM}, we have $-\partial_{\e} f_{0}(\e) =\delta(\e-\e_F)$, such that functions $\mathcal{F}_{\Theta}$, $\mathcal{F}_{\Lambda}$ are respectively reduced to $\mathcal{F}_{\Theta}=\Theta(\e_F), ~\mathcal{F}_{\Lambda} = \Lambda(\e_F)$. 

We want to mention that $\Theta(\e)$, as plotted in Fig.~\ref{fig:fig2}(b), is in effect the density of states (DOSs) for a single Weyl cone at energy $\e$, which can be modulated by the presence of different fields. While $\Lambda (\e)$ (i.e., $\mathcal{F}_{\Lambda}$) presents evident oscillations even with much bigger amplitudes than that of $\Theta (\e)$~\cite{SM}, the manifested oscillations of $\bm{j}_z$ will be mainly determined by $\Theta(\e)$. This is because of the large ratio of scattering time scales $\tau_v/\tau_a$ in WSMs. For instance, based on $\tau_v/\tau_a \sim (2k_0/k_F)^4$ with $2k_0$ the Weyl node separation and $k_F$ the Fermi wave vector~\cite{scattering_ratio_Pesin_2014prx}, this ratio can be estimated as $\tau_v/\tau_a \sim 10^4$ in WSM Na$_{3}$Bi~\cite{exp_scattering_na3bi_2014sci}. %More discussions can be found in Ref.~\cite{SM}. 

After some algebra with Eqs.~(\ref{eq:deltamu}, \ref{eq:currents}), the CA-induced response can be found to be $\bm{j}^C_z \propto B\mu^C_s$, where $\mu^C_s = s \tau_v eE_{z} v_F/\Theta$ is the CCP, chirality-dependent as expected for each Weyl node. Straightforwardly, a charge current linear in $E_z$ is obtainable. The magnetic field dependency of $j^C_z$ is revealed by Fig.~\ref{fig:fig2}(c), where a quadratic-in-$\bm{B}$ (inset) and linear-in-$\bm{B}$ signature is observed in the weak and strong field limit, respectively. This directly recovers the conventional result as discussed for the linear magneto-transports in earlier studies~\cite{spivak_LMC_2013prb,burkov_magnetotransport_2014prb,Burkov_phe_2017prb}, and more relevant discussions %for the linear magneto transports in different field regimes 
can be found in SM~\cite{SM}. Next, we will investigate the CA-induced NPEs. 
\begin{comment}
Note that, the amplitudes of $\Theta(\e)$ become infinitely large at the Van Hove singularities $\e_F =\sqrt{2|n|\alpha (1-\beta^{'2})(\hbar\omega_c)^2}$, rendering $\Theta$ oscillate with $B$ in a certain period [also see Fig.~\ref{fig:fig4}(a)]. Note that $\beta^{'2} =\beta^2 +\beta^2_x$, different from $\beta^2$. 

\textcolor{red}{This part should be for the linear transport}

%\begin{equation}
%    \begin{split}
%        \mathcal{F}_{\Theta}=\Theta(\e_F),
%        ~~\mathcal{F}_{\Lambda} = \Lambda(\e_F). \label{eq:reducedFs}
%    \end{split}
%\end{equation}
In the absence of any transverse electric field, 

As a result, CCP in Eq.~(\ref{eq:deltamu}) is reduced to $\mu^C_s = s \tau_v eE_{z} v_F/\Theta$ (i.e., $\Theta^{-1} \propto \mu^C_s$), which is chirality-dependent as expected for each Weyl node, and the CA-related current is then found as $\bm{j}^C_z \propto sB\mu^C_s$. 
\end{comment}

%%%%%%%%%%%%%%%%%%%%%%%%%%%%%%%%%%%%%%%%%%%%%%%%%%%%%%%%%
\begin{figure}[tp!]
	\begin{center}
		\includegraphics[width=0.465\textwidth]{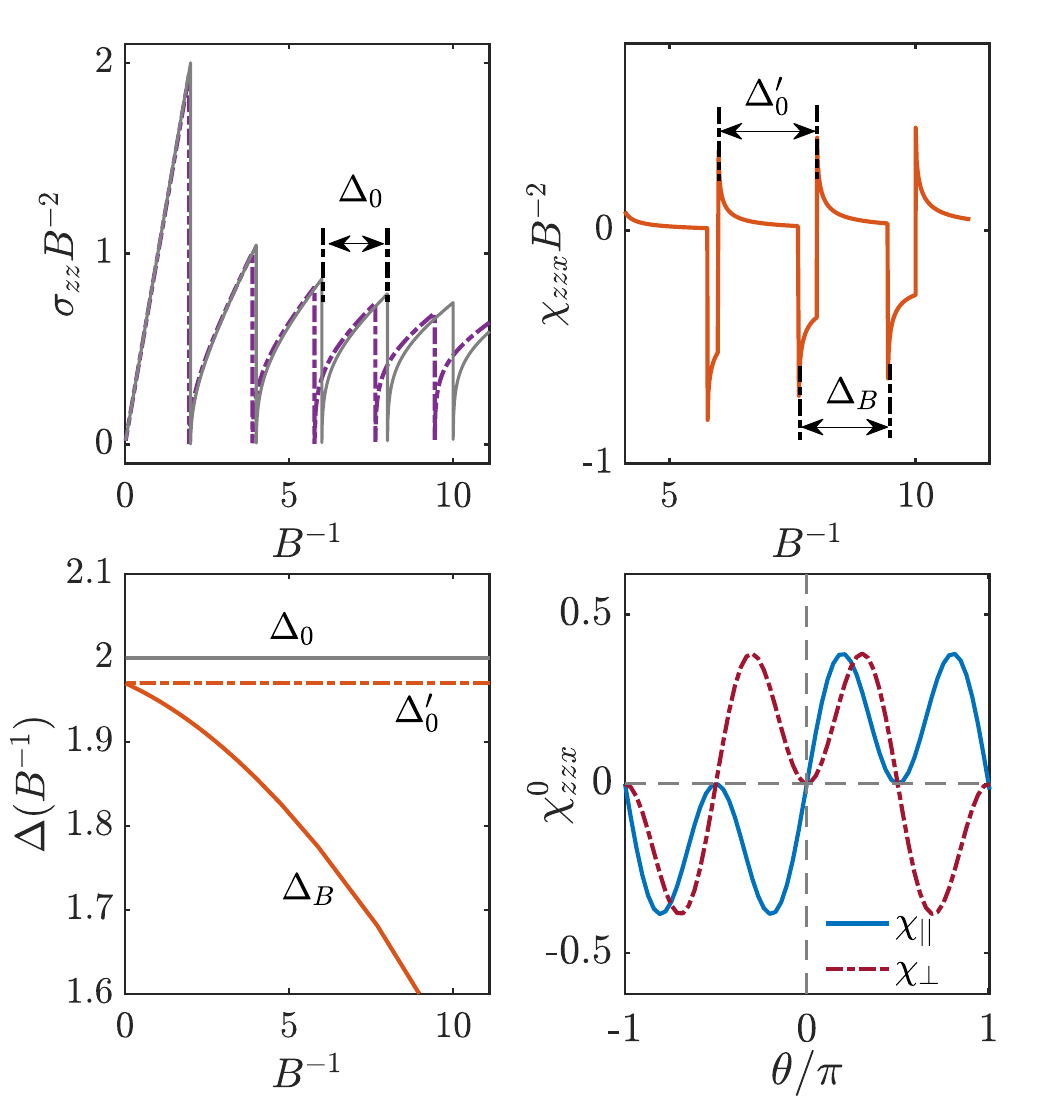}
		\llap{\parbox[b]{162mm}{\large\textbf{(a)}\\\rule{0ex}{81mm}}}
		\llap{\parbox[b]{82mm}{\large\textbf{(b)}\\\rule{0ex}{80mm}}}
		\llap{\parbox[b]{166mm}{\large\textbf{(c)}\\\rule{0ex}{42mm}}}
		\llap{\parbox[b]{84mm}{\large\textbf{(d)}\\\rule{0ex}{42mm}}}
	\end{center}
	\vspace{-7mm}
	\caption{(Color online) Periodic oscillations in $B^{-1}$ of the linear ($\sigma_{zz}$) and nonlinear conductivity ($\chi_{zzx}$) in panel (a) and (b), respectively. Note that the additional $\beta^E_{x}$ does not make big difference on the magnitudes of $\sigma_{zz}$. The constant periods $\Delta_0, \Delta^{\prime}_0$ and the magnetic field dependent $\Delta_B$ are compared in panel (c). Here, $E_F =v_F \sqrt{e\hbar} (=1 \hbar \omega_c\big|_{B=1})$ is taken. Panel (d) presents the angular dependencies of the longitudinal ($\chi_{||}$) and transverse ($\chi_{\perp}$) components of the magneto conductivity tensor, here $\theta= \<\bm{E, B}\>$ as shown in Fig.~\ref{fig:fig1}.} 	\label{fig:fig3}
	\vspace{-2mm}
\end{figure}
%%%%%%%%%%%%%%%%%%%%%%%%%%%%%%%%%%%%%%%%%%%%%%%%%%% FIG 3
\textcolor{blue}{\it{Chiral anomaly-induced nonlinear transports:}} As shown in Fig.~\ref{fig:fig2}(b) and (c), the DOSs $\Theta$ (or $\Theta^{-1}$) get modified when $\beta^E_x \neq 0$ or/and $\beta^w_y \neq 0$ compared to the non-perturbed case. Such modification turns out to be responsible for the nonlinear CA-related responses. In effect, %as shown in Fig.~\ref{fig:fig1}(c1), %in the geometry of co-planar fields with $\bm{E\cdot B} \neq 0$ lying in the $x-z$ plane,
the transverse component $E_x \hat{\bm{x}}$ (as $\beta^E_x \neq 0$) can bring in additional $E$-dependence for the generated CCP in WSMs besides the longitudinal $E_z \bm{\hat{z}}$, making the CA-induced nonlinear responses possible. Considering this, we can formulate the charge current $\bm{j}_z$ as $\bm{j}_z = \sigma_{zz}E_{z} + \chi_{zzx} E_{z}E_{x} + \dots$, where $\sigma_{zz}$ and $\chi_{zzx}$ are the linear and second-order nonlinear conductivity tensors, respectively. Based on Eq.~(\ref{eq:currents}), it is clear that $\sigma_{zz}$ and $\chi_{zzx}$ are fully determined by $\Theta(\e), \Lambda(\e)$. 
%In the linear transport regime, the longitudinal $E_z \bm{\hat{z}}$ leads to the CA-related linear electric transports. %as mainly determined by the nonzero $\bm{E\cdot B}$,
%well captured by a series of conductivity tensors, i.e., $\bm{j}_z = \sigma_{zz}E_{z} + \chi_{zzx} E_{z}E_{x} + \dots$, where $\sigma_{zz}$ and $\chi_{zzx}$ are fully determined by functions $\Theta(\e), \Lambda(\e)$. 
%when rewritten into linear and nonlinear (up to $2^{nd}$ order) components. 

Irrespective of the particular magnetic field regime, it is allowed to expand $\Theta^{-1}$ in terms of $\beta^{E}_x$ perturbatively to study the effect of $\beta^E_x$, which can be given as, 
\begin{equation}
    \Theta^{-1}(\e) =\Theta^{-1}_0(\e) + \beta^{E}_{x}\Theta^{-1}_r(\e) + \mathcal{O}[(\beta^{E}_x)^2] .   \label{eq:rreducedFs}
\end{equation}
The above expansion is valid as long as $\beta^E_x \ll 1$ is satisfied (similarly, for $\Lambda(\e)$ we have $\Lambda_0, \Lambda_r$~\cite{SM}). As such, $\Theta^{-1}_0, \Theta^{-1}_r$ are $E_x$-independent. The modifications denoted as $\delta \Theta^{-1} =\Theta^{-1}-\Theta^{-1}_0$ (similarly $\delta \Lambda$), as expected from Eq.~(\ref{eq:rreducedFs}), are found to be roughly linear in $\beta^E_x$ under various magnetic fields at the limit of $\beta^E_x \ll 1$, which now are given in Fig.~\ref{fig:fig2}(d). Additionally, a linear dependency of $\delta \Theta^{-1}$ in $\bm{B}$ is also observed in the weak field regime (black solid line, inset). 

Recapping $\mu^C_s \propto E_z \Theta^{-1}$ and $\bm{j}_z \propto BE_z \Theta^{-1}$, one obtains 
\begin{equation}
    \mu^C_s(\e) = \mu^C_{s,0}(\e) + \delta \mu^C_s(\e).
\end{equation}
Here $\mu^C_{s,0}(\e)\propto E_z \Theta^{-1}_0$ denotes the conventional CCP that governs the linear magneto transports in WSMs, i.e. $\sigma_{zz} \propto \mu^C_{s,0} \propto \Theta^{-1}_0$. On the other hand, we also have $\delta \mu^C_s(\e) \propto E_z \beta^E_x \Theta^{-1}_r$ for the the deviation of CCP, which exactly gives rise to the second-order nonlinear responses. A comparison between $\mu^C_s$ and $\delta \mu^C_s$ is schematically shown in Fig.~\ref{fig:fig1}(c2). Based on the above analysis, the nonlinear conductivity $\chi_{zzx}$ can be obtained as
\begin{equation}
    \chi_{zzx} = c_0 \big[(\tau_v-\tau_a)\Theta^{-1}_r(\e) + \tau_a \Lambda_r(\e)\big].  \label{eq:nonlinear_qs}
\end{equation}
Hereafter, $\sum_s$ is omitted for simplicity. Note that, all the field dependencies of the conductivity will be solely revealed by functions $\Theta^{-1}_r, \Lambda_r$, and $\Theta^{-1}_r$ explicitly appears in $\delta \mu^C_s$, connecting with the CA in WSMs unambiguously.  
%Similarly, the oscillations of $\chi_{zzx}$ is also mainly determined by (CA-manifested) $\Theta^{-1}_{r}$ because of $\tau_v\gg \tau_a$. 

Moving to the semiclassical regime with $\e_F \gg \hbar \omega_c$ (i.e., weak magnetic field regime) where the LL description is not valid, the CA-induced and the normal-drift-contributed nonlinear conductivity can be obtained based on Eq.~(\ref{eq:nonlinear_qs}) as~\cite{SM} 
\begin{equation}
    \chi^{C}_{zzx}   \propto w^{s}_y \tau_v B/\Tilde{\e}^{2}_s, \quad 
    \chi^{N}_{zzx} \propto w^{s}_y \tau_a \Tilde{\e}^2_s/B
\end{equation}
respectively. Here $\Tilde{\e}_s = (\e-s Q_0)/v_F\hbar$, $w^s_y = s w_y /v_F$, and approximations $|\beta^E_x|^2 \ll 1, |\beta^{w}_{y}|^2 \ll1$ have been applied in the derivations. Note that, the different parameter dependencies between $\chi^C_{zzx}$ and $\chi^N_{zzx}$ are helpful to distinguish their contribution to the nonlinear response, though typically the former i.e. CA, will be the leading contributor in WSMs due to $\tau_v/\tau_a \gg 1$. It is also interesting to notice, that the CA-induced nonlinear conductivity tensor in the semiclassical regime is linear in magnetic field, in contrast to that of the linear transports. This also makes the CA-induced nonlinear response easily distinguished from other possible plaguing effects~\cite{Liang_currentjet_2018prx,Yang-currentjet_2019prb,Dos_currentjet_2016NJPh,Arnold_currentjet_2016Nc}.
%We want to mention that, the above results are consistent with that found in early studies where the nonlinear conductivity is obtained using Boltzmann transport theory without involving any LLs~\cite{Li_caNLH_2021prb,Nandy_caNLH_2021prb,Zeng_caNTH_2021}. 

Similarly, the conductivity tensor in the ultraquantum regime with $\hbar \omega_c\gg \e_F$ can be obtained as
\begin{equation}
   \chi^{C}_{zzx} \propto w^{s}_y  \tau_v,\quad \chi_{zzx} \approx \chi^{C}_{zzx}. \label{eq:ultra_nonlinear_reduced} % \frac{w^{s}_y e^3 \tau_v }{2 \pi^2 \hbar^2 v_F}
\end{equation}
In this regime, only the chiral zeroth-LL is occupied due to the large LL spacing under the strong magnetic field. Interestingly, $\chi_{zzx}$ is magnetic field-independent as well as Fermi energy-independent in this regime, different from that in the semiclassical limit. As a result, the CA-induced nonlinear responses can survive under the strong magnetic field only for the pair of achirally tilted Weyl nodes ($w^{+}_{y}\neq -w^{-}_y$). 

Note that, the nonlinear conductivity discussed above explicitly depends on the band tilt, thus clearly implying a joint effect of the transverse electric field and the band tilt. Specifically, the nonlinear current is given by $\bm{j}_z \propto \tau_v w^{s}_y E_{x} E_{z} B$, which can be effectively provided by $(\bm{E\cdot B})(\bm{E\times w})$, exactly the configuration employed for the CA-induced nonlinear transport effects in recent works~\cite{Li_caNLH_2021prb,Nandy_caNLH_2021prb,Zeng_caNTH_2021}.

\textcolor{blue}{\it{Signatures for experimental probing:}} It is now known that quantum oscillation effects origin from the successive crossing of the Fermi energy by LLs, which manifests as the periodic change in the electron or hole DOSs~\cite{book_qoe}. As shown in Fig.~\ref{fig:fig2}(b), the amplitudes of $\Theta^{-1}(\e)$ drops down to zero periodically, attributing to the Van Hove singularities in $\Theta(\e)$ at $\e_F =\sqrt{2|n|\alpha (1-\beta^{'2})(\hbar\omega_c)^2}$ (i.e. infinitely large DOSs, see in Ref.~\cite{SM}). This feature is also present in CCP as well as the magneto conductivity in WSMs, for which the oscillating period can be obtained as
\begin{equation}
    \Delta(B^{-1}) = 2 \alpha(1-\beta^{'2}) (v_F\hbar/\e_F)^2 e/\hbar.
\end{equation}
Here $\beta^{'}=\sqrt{\beta^2 + (\beta^{w}_z)^2}$ with $\beta^{w}_z = s w_z/v_F$ accounting the band tilt along $\bm{z}$-direction. As the value of $w_z$ only changes the relative magnitudes of the oscillating quantities, here for simplicity we consider the case of $w_y\neq 0, w_z=0$. Thus, the period is simply given as $\Delta(B^{-1}) =2 \alpha^3 (v_F\hbar/\e_F)^2 e/\hbar$, which obtains modification from factor $\alpha^3$. %To better present the periodic oscillation,  in Fig.~\ref{fig:fig3}(a). 

As shown in Fig.~\ref{fig:fig3}(a), $\sigma_{zz}B^{-2}$ is plotted as a function of $B^{-1}$, showing that a finite $\beta^E_x$ leads to a decreasing oscillating period (dashed line), in contrast to the case of $\beta^E_x=0$ with the period being a constant (grey solid, $\Delta_0$). %This feature is consistent with earlier studies. 
Interestingly, since the second-order nonlinear transports link with $\delta \mu^C_s$, which as discussed earlier is proportional to $(\Theta^{-1}-\Theta^{-1}_0)$, the corresponding conductivity $\chi_{zzx}B^{-2}$ can reveal \textit{two different oscillating periods} simultaneously. As shown in Fig.~\ref{fig:fig3}(b), one period remains as a constant $\Delta_{0}^{\prime}$ due to $\Theta^{-1}_0$, while the other one $\Delta_B$ corresponding to $\Theta^{-1}$ exhibits an evident magnetic field dependency. The above-mentioned oscillation periods are shown in Fig.~\ref{fig:fig3}(c) for comparison. Note that, the difference between $\Delta_0$ and $\Delta_{0}^{\prime}$ depends on the exact value of $\beta^{w}_y$, and the decreasing period appears as long as the transverse electric field ($\beta^{E}_x \neq 0$) is present. We want to point out that, such a remarkable oscillating feature, different from all the known quantum oscillation effects, is characteristic of the CA-induced second-order nonlinear transports.

Besides the oscillation periods, angular dependence is another probable signature that is important in studying the CA-related transports in WSMs. Though the longitudinal charge currents (and conductivity tensors) are written to be locked to $\bm{B}=B\bm{\hat{z}}$, it naturally gives risse to the planar transport components when expressed in terms of the crystal axes or the orthogonal directions with respect to $\bm{E}$ ($\<\bm{E, B}\> =\theta$, see Fig.~\ref{fig:fig1}). As has been discussed for the linear transports~\cite{Deng_qc_2019prb,kamal_cas_2020prr}, one can obtain $\sigma_{||} =\sigma_{zz} \cos^2{\theta}, \sigma_{\perp} = \sigma_{zz} \cos{\theta}\sin{\theta}$ as the LMC and PHE conductivity tensor respectively. 
%This angular dependence, however, was found closer to $\cos^4{\theta}$ instead of $\cos^2{\theta}$ for LMC in experiments~\cite{2015Sci_CA_exp,2016Nc_CA_exp,Liang_reviewpaper_2021}. In a recent study, it shows that the angular dependence in the weak magnetic field limit should be replaced by the stronger $\cos^6{\theta}$ for LMC ($\cos^5{\theta}\sin{\theta}$ for PHE) due to the modulation factor~\cite{Deng_qc_2019prb}. 

In the nonlinear transports, we have $\chi_{||}=\chi_{zzx} \cos^2{\theta}\sin{\theta}$ and $\chi_{\perp} =\chi_{zzx}\sin^2{\theta}\cos{\theta}$, which can be taken as CA-induced nonlinear LMC and PHE tensors respectively. The angular dependencies for $\chi_{||}, \chi_{\perp}$ are now plotted in Fig.~\ref{fig:fig3}(d). Interestingly, all the NPE conductivity tensors switch signs when changing the angle $\theta$, and the nonlinear LMC is odd while the PHE is even in $\theta$. These features of NPEs are fundamentally different from that in the linear transport regime, hence are helpful for the probing of the CA-induced NPEs in experiments. %Note that, the angular dependence in the linear regime was found closer to $\cos^4{\theta}$ instead of $\cos^2{\theta}$ for LMC in experiments~\cite{2015Sci_CA_exp,2016Nc_CA_exp,Liang_reviewpaper_2021}, which in the later theoretical work has been interpreted to be replaced by the stronger $\cos^6{\theta}$ for LMC ($\cos^5{\theta}\sin{\theta}$ for PHE) due to the modulation factor~\cite{Deng_qc_2019prb}. 
The angular dependencies of the NPEs are also expected to be free of any squeezing effects and thus robust for experimental probing, in contrast to the linear case where a closer angular dependency $\cos^4{\theta}$~\cite{2015Sci_CA_exp,2016Nc_CA_exp,Liang_reviewpaper_2021} or $\cos^6{\theta}$~\cite{Deng_qc_2019prb} was found for LMC instead of the predicted $\cos^2{\theta}$. Additionally, the CA-induced NPEs discussed in this work are measurable through frequency lock-in measurement using AC current, as has been conducted successfully in experiments for other nonlinear effects~\cite{nhe_kang_2019NatMa,nhe_ma_2019Natur,third_nhe_Yang_2021NatN}. 

%rendering $\Theta$ oscillate with $B$ in a certain period (and periodically in $1/B$).

\textcolor{blue}{\it Conclusion:} By considering the configuration with non-collinear electric and magnetic field and simultaneously involving a band tilt, we show that, besides the fascinating angular dependence, the CA-induced NPEs for WSMs display unique quantum oscillation features in the intermediate magnetic field regime. Specifically, in striking contrast to the oscillations with a single period in $B^{-1}$ obtained for the positive LMC and PHE in the linear transport regime~\cite{Deng_qc_2019prl,Deng_qc_2019prb,Deng_qc_2019prr,
Shao_lorentz_wsm_2021iop,kamal_cas_2020prr}, the CA-induced NPEs exhibit quantum oscillations with two different period scales, one remaining constant while the other decreasing with increasing $B^{-1}$. We show that such behaviors are directly connected to chiral anomaly through the deviation in the chiral chemical potential, thereby making the characteristic quantum oscillation signatures in the non-linear and linear response regimes uniquely suitable for identifying CA in WSMs.

\textcolor{blue}{\it Acknowledgement:} The work is supported by the National Key R\&D Program of China (Grant No. 2020YFA0308800), the NSF of China (Grants Nos. 12104043, 11734003, 12061131002)), the Strategic Priority Research Program of Chinese Academy of Sciences (Grant No. XDB30000000), and the fellowship of the China Postdoctoral Science Foundation (Grant No. 2021M690409). S. N. acknowledges the National Science Foundation Grant No. DMR-1853048. S. T. thanks the ARO Grant No. W911NF-16-1-0182 and Grant No. NSF2014157 for support.  C. Z. also thanks J.M. Shao for the useful discussions.

\bibliography{my}

\end{document}